%% file: WPC-Corr.tex
\documentclass[runningheads]{llncs}
\usepackage{multirow}
\usepackage{makecell}
\usepackage{caption}
\usepackage{subcaption}
\usepackage{graphicx}

\title{WPC: Whole-picture Workload Characterization}
\author{
Lei Wang\inst{1,2,3} \and
Kaiyong Yang\inst{1,2} \and
Chenxi Wang\inst{1,2} \and
Wanling Gao\inst{1,2,3} \and
Chunjie Luo\inst{1,2,3} \and
Fan Zhang\inst{1,2} \and
Zhongxin Ge\inst{1,2} \and
Li Zhang\inst{1,2} \and
Guoxin Kang \inst{1,2} \and
Jianfeng Zhan\inst{1,2,3}\thanks{Jianfeng Zhan is the corresponding author.}
}
\institute{Institute of Computing Technology, Chinese Academy of Sciences, Beijing 100190, China
\email{\{wanglei\_2011,zhanjianfeng\}@ict.ac.cn} \and University of Chinese Academy of Sciences \and International Open Benchmark Council (BenchCouncil)\\
}

\UseRawInputEncoding
\begin{document}
\maketitle
\thispagestyle{plain}
\pagestyle{plain}


\begin{abstract}


This article raises an important and challenging workload characterization issue: can we uncover each critical component across the stacks contributing what percentages to any specific bottleneck? The typical critical components include languages, programming frameworks,  runtime environments,  instruction set architectures (ISA), operating systems (OS), and microarchitecture. Tackling
this issue could help propose a systematic methodology to guide the software and
hardware co-design and critical component optimizations.

We propose a whole-picture workload characterization (WPC) methodology to answer the above issue. In essence, WPC is an iterative ORFE loop consisting of four steps: \underline{O}bservation, \underline{R}eference, \underline{F}usion, and \underline{E}xploration. WPC  observes different level data (observation), fuses and normalizes the performance data (fusion) with respect to the well-designed standard reference workloads suite (reference), and explores the software and hardware co-design space (exploration) to investigate the impacts of critical components across the stacks. We build and open-source the WPC tool. Our evaluations confirm WPC can quantitatively reveal the contributions of the language, framework, runtime environment, ISA, OS, and microarchitecture to the primary pipeline efficiency.



\end{abstract}
\section{Introduction}\label{sec:introduction}

The hardware renaissance has witnessed a 50,000-fold performance improvement since 1978~\cite{hennessy2019new}. In this background, high-productivity languages like Java and Python with complex programming frameworks like Hadoop and TensorFlow are gaining popularity over performance-oriented languages like C and C++. Workload characterization is to understand the characteristics and behaviors of the diverse workloads, playing a significant role in the design, optimization, and evaluation of systems and architectures~\cite{calzarossa2016workload}. With respect to the performance-oriented workloads, the workloads programming with high-productivity languages
have many heavier stacks, consisting of languages, programming frameworks (e.g., MapReduce or MPI),  runtime environments (e.g., JVM, HDFS file systems, and Yarn scheduler),  instruction set architectures (ISAs), operating system (OS), microarchitecture and other critical components, which raises serious challenges to workload characterization. To facilitate
exploring the software and hardware co-design space, we need to understand the impact of each component across the stacks on any specific bottleneck.

This article raises an important and challenging workload characterization issue: can we uncover each critical component across the stacks contributing what percentages to any specific bottleneck? A bottleneck may be a pipeline bottleneck (microarchitecture), storage or network, or other significant ones. If yes, the answer will provide a hint on how to explore software and hardware co-design space.
Unfortunately, no previous work can answer this issue. For example, the state-of-the-practice tool,  VTune, can perform a top-down performance analysis to uncover pipeline bottlenecks of out-of-order execution processors on a specific processor. Still, it can not reveal the impacts of those critical components on the pipeline bottlenecks in a quantitative manner -- who contributes what percentages of a specific pipeline bottleneck?

Previous methodologies only perform workload characterization on a single level like intermediate representation (IR), ISA, or
microarchitecture~\cite{6557175,panda2018wait,parsec,yasin2014deep}. They can not answer the above issue.
For example, Yasin et al. uncover pipeline bottlenecks of Intel processors with the top-down performance analysis at a specific microarchitecture~\cite{yasin2014deep}.
It takes the instruction streams on a specific microarchitecture as the input to perform analysis; however, it can not quantify the overheads of critical components of the software and hardware stacks. Also, a single IR or ISA level analysis takes either intermediate representation or binary stream as the input, omitting the stacks' ISA level or microarchitecture level characteristics accordingly.

One intuition is integrating the existing single-level workload characterization tools such as LLVM (IR level), Pin (ISA level), or Perf (microarchitecture level). Someone may argue that a simple integration of three-level data can obtain more insights across the stacks (more details in Section~\ref{MovivationSubSection}). Unfortunately,  the same characteristics at each level of the system stacks have different metrics or observation value ranges, and the observation tools also distort the results of the observations. For example, the IR-level tool can not obtain the characteristics of the third-party libraries, and the ISA-level tool can not capture the OS systems function calls. So only naively integrating three-level data can not naturally lead to new insights and reveal each component's quantified effect on the overheads of the workload across stacks.

\begin{figure}[h]
\centering
\includegraphics[scale=0.5]{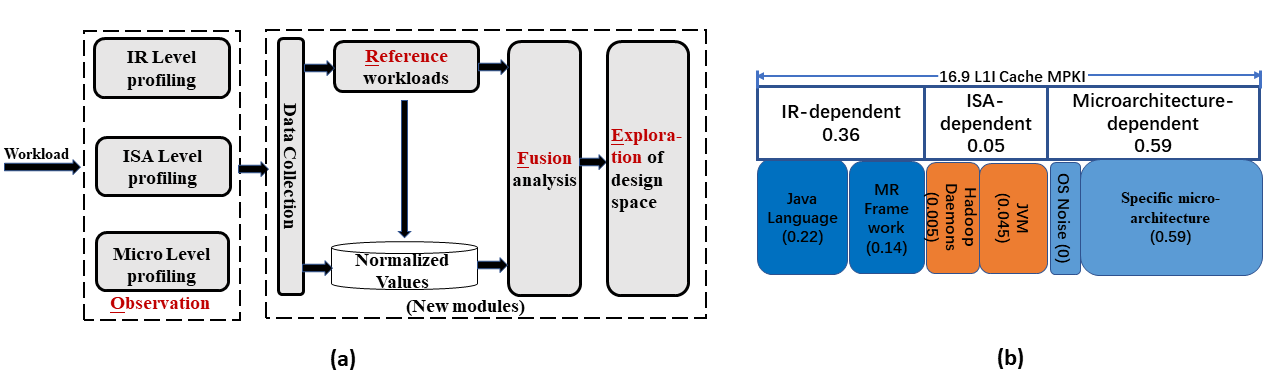}
\caption{Overview of the WPC methodology. Figure 1-a outlines the WPC methodology and tool. In a case study, Figure 1-b demonstrates that WPC can break down the impacts of critical components on the L1I Cache Misses of the Bayes-Hadoop workload. The components include Java language, MapReduce programming framework, Hadoop daemons, JVM, OS noise, and the specific microarchitecture (Intel Xeon Gold 5120T). The details of the case study are in Section~\ref{EvaluationsScaleOut}.}
\label{method1fig301Intro}
\end{figure}

As shown in Fig.~\ref{method1fig301Intro}, this paper presents a whole-picture workload characterization (in short, WPC) methodology and implements the WPC tool. The name of WPC originates from the fact that it can uncover each critical component across the stacks, contributing to how many percentages to the pipeline efficiency.  In essence,  WPC is an iterative ORFE loop consisting of four steps: \underline{O}bservation, \underline{R}eference, \underline{F}usion, and \underline{E}xploration. The observation across the stacks is the first step.
Currently, WPC observes workload characterizations at three primary levels, including IR, ISA, and microarchitecture.
Second, WPC proposes a suite of standard reference workloads. The standard reference workloads are well-designed, simple workloads that accurately control instruction, data, and branch localities. Their behaviors are deterministic and consistent, and the little deviation of behaviors across different stacks is explainable. With the standard reference workloads, we quantify the characteristic variations of the complex workloads across the system stacks. The third step is fusion. We fuse the analysis results from different observations through statistical analysis and standard reference workloads-based analysis. Furthermore, we use the normalized impact factors to reflect the contributions of various components across the stacks to the bottlenecks.
The fourth step is exploring the software and hardware co-design space and further investigating the impacts of critical components.

As the first step, this article focuses on pipeline bottlenecks, e.g., high L1I Cache Misses. We leave the other bottlenecks open and future work. WPC can currently analyze instruction, data, and branch localities. We will extend WPC to other bottlenecks in the near future.  WPC can apply to all single-thread or multi-thread C or Java workloads running on the X86 and ARM systems. The WPC tool is convenient to use, and most analyzed data reported in the paper are automatically generated by the WPC tool. The current WPC implementation also has limitations, which we will extend soon. The ISA-level tools lack kernel visibility and can only capture user-level code. We will add the OS-level profiling based on Systemtap. WPC only supports the microarchitectural workload characterization on the actual system stacks, and we will support the analytical model system~\cite{7442561} and the simulation/emulation system~\cite{6322869}. It does not yet support virtualization systems.

Our contributions are as follows.

1) We raise an important and challenging workload characterization issue: can we propose a systematic methodology to uncover each critical component across the stacks contributing to how many percentages to a specific bottleneck?

2) We reveal that performing single-level workload characterization alone will lead to contradicted conclusions. Also, We uncover a simple integration of three-level data can not naturally lead to new insights.

3) We propose the WPC methodology and build the open-source WPC tool to help answer the above issue.

4) Our experiments show WPC can quantify and break down the effect of each component across stacks on the overheads of the workload. We reveal that for the L1I Cache Misses of the Bayes-Hadoop workload, the normalized impact factor of the Java language, the MapReduce programming framework, the Hadoop daemons, the JVM runtime environment, and the specific microarchitecture (Intel Xeon Gold 5120T) is 0.22, 0.14, 0.005, 0.045, and 0.59, respectively; this conclusion helps us improve the performance by exploring the co-design space.

The remainder of the paper is organized as follows. Section II explains our motivations. Section III presents the WPC methodology and tool. Section IV is a comprehensive evaluation. Section V summarizes the related work. Section VI concludes.

\section{Background and Motivation}
\subsection{Background}

The scale-out data analytical workloads, which refer to data-level or request-level workloads often developed using high-productivity languages based on a distributed framework like Hadoop, are typical modern  workloads~\cite{ferdman2011clearing,barroso2018datacenter}.
We take the Bayes-Hadoop workload as an example to illustrate its execution process. Each workload execution consists of seven layers: languages (Java), programming frameworks (MapReduce), framework daemons (Hadoop daemons), runtime environments (JVM), ISA, OS, and microarchitecture.
The scale-out Hadoop workload's source codes are implemented with Hadoop's MapReduce programming framework in Java. The Javac compiler compiled the scale-out Hadoop workload's source codes into bytecodes (IR codes) and packed them as the Jar file.
MapReduce programming framework and the Java language are related to IR level. The Hadoop daemons and JVM runtime environment are related to ISA level. As a distributed framework, the Hadoop daemons, including the HDFS filesystem, the Yarn scheduler, and other management daemons, run above the JVM runtime environment. The Jar file is submitted to the Hadoop daemons and runs as the Jar file on the JVM. The JVM interprets byte codes to machine codes specific to different ISAs and manages the Java program. The machine codes are instantiated as the OS process, running on the processor as the instruction stream dependent on a specific microarchitecture.

Likewise, for the MPI workloads implemented with C, the MPI workload's source codes are compiled to the IR by Clang. Its runtime environment includes a set of libraries, such as the MPI library and the LIBC library, with which the IRs are linked to the machine codes. Then the binary machine code of the workload runs as the OS process. Finally, the workload runs on the processor as the instruction stream.

\subsection{Motivations}\label{MovivationSubSection}

We take the state-of-the-practice and state-of-the-art workload characterization as motivating examples to demonstrate why we propose the WPC methodology.

\subsubsection{Workloads and Evaluation methodology}~\label{MotivationMetrics}

SPEC CPU2017 is the state-of-the-practice CPU benchmark suite~\cite{panda2018wait}, written in performance-oriented languages like C, C++, and Fortran. We evaluate a total of 23 throughput-oriented SPECrate workloads from SPEC CPU2017.  We choose eight typical scale-out workloads implemented with Hadoop frameworks, i.e.,  Bayes-Hadoop, Pagerank-Hadoop,  Kmeans-Hadoop, CF-Hadoop, CC-Hadoop, Sort-Hadoop, Grep-Hadoop, and MD5-Hadoop. These scale-out Hadoop workloads are included in three influential benchmarks suites, CloudSuite~\cite{ferdman2011clearing}, BigDataBench~\cite{BigDataBench_hpca}, and HiBench~\cite{hibench}.

We choose instruction reuse distance (in short, Reuse\_Dist)~\cite{1628963}, instruction Reuse\_Dist, and L1I Cache misses per kilo instructions (MPKI) as metrics to measure the instruction localities at the IR, ISA, and microarchitecture levels, respectively. Among them, the instruction Reuse\_Dist metric calculates the average distance between two consecutive accesses to the same instruction address. A larger instruction Reuse\_Dist value implies a lower locality. The metric of instruction Reuse\_Dist is calculated at the IR and ISA levels. To measure instruction Reuse\_Dist, we use LLVM and Hotspot at the IR level and Pin and DynamoRIO at the ISA level. We use Perf to measure the L1I Cache MPKI at the specific microarchitecture.

\subsubsection{Performing single-level workload characterization alone will lead to contradicted conclusions}~\label{Motivation_single}

Fig.~\ref{motivationfig0001} reports the instruction localities at the IR, ISA, and microarchitecture levels. The evaluation platform is the Intel Xeon Gold 5120T. The metrics are instruction Reuse\_Dist (the central axis) at the IR and ISA levels and  L1I Cache MPKI at the microarchitecture level (the secondary axis). We have the following findings:

The observations significantly differ at three levels, and performing single-level workload characterization alone will lead to contradicted conclusions. From the IR level, the SPECrate workloads and the scale-out Hadoop workloads have similar instruction Reuse\_Dist, and the average gap is no more than 1.2 times. From the ISA level, the average instruction Reuse\_Dist of the SPECrate workloads is 9.2 times that of scale-out Hadoop workloads, which implies that the instruction locality of the SPECrate workloads is worse than that of the scale-out Hadoop workloads. However, from the microarchitecture level, the average L1I Cache MPKI of the SPECrate workloads is only 0.4 times that of the scale-out Hadoop workloads, which implies that the instruction locality of the SPECrate workloads is better than that of the scale-out Hadoop workloads, and this corroborates the previous observations ~\cite{ferdman2011clearing,10.1145/2749469.2750392,7920850}. The IR-level observation implies SPECrate,  and the Hadoop workloads have similar instruction locality. However,  the ISA-level observation suggests SPECrate has a worse instruction locality. Instead, the microarchitecture observation indicates the Hadoop workloads have worse instruction locality. Those observations contradict each other.

\subsubsection{Simple integration of  three-level profiling data does not help lots}~\label{Motivation_integration}

We integrate three-level profiling data to find more insights. To understand the relationship of observation characteristics among the IR, ISA, and microarchitecture levels, we adopt the Pearson correlation coefficient~\cite{1996Pearson} to validate the correlation. Pearson correlation coefficient ranges from -1 to 1, and its absolute value shows the correlation. The larger the absolute value is, the stronger the correlation is between two variables. A positive value means a positive correlation and vice versa.

\begin{figure}[h]
\centering
\includegraphics[scale=0.3]{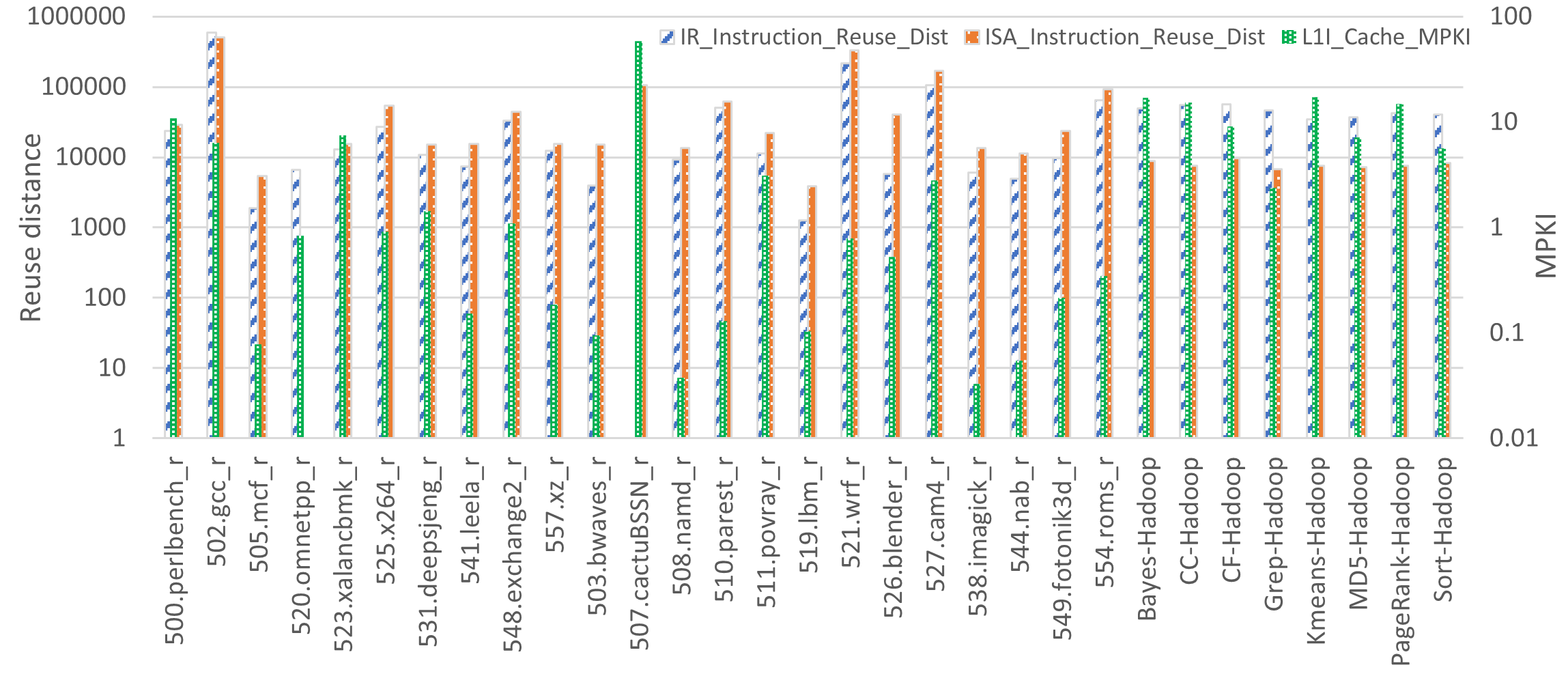}
\caption{This figure reports the raw data of the instruction localities at the IR, ISA, and microarchitecture levels. Our analytics on the raw data show: that (a) performing single-level workload characterization independently at three levels will lead to contradicted conclusions. (b) Simple integration of three-level profiling data does not help lots. The same characteristics at each level have different metrics or observation value ranges, and the tools at each level also distort the results of the observations. For those reasons, drawing any critical conclusion or insight isn't straightforward.}
\label{motivationfig0001}
\end{figure}

We still analyze the instruction localities at three levels, and the metrics are the same, like~\ref{Motivation_single}.
From Fig.~\ref{motivationfig0001}, the SPECrate workloads' instruction Reuse\_Dist have similarities at the IR and ISA levels, and the correlation coefficient between the IR  and ISA levels is 0.97, which implies a strong positive correlation. We also note that the instruction Reuse\_Dist of the scale-out Hadoop workloads at the ISA level is significantly less than those at the IR level, and the correlation coefficient between them is 0.52. So, the scale-out Hadoop workloads' instruction Reuse\_Dist have no significant similarities across the IR and ISA levels.
The instruction Reuse\_Dist of SPECrate workloads at the ISA level does not have significant similarities with the L1I Cache MPKI at the microarchitecture level, and the correlation coefficient is only 0.12. The instruction Reuse\_Dist of scale-out Hadoop workloads at the ISA level also do not have significant similarities with the L1I Cache MPKI at the microarchitecture level, and the correlation coefficient is only 0.15.

From the above analysis, it isn't straightforward to draw any critical conclusion or insight for a reason we have discussed in Section~\ref{sec:introduction}. We repeat here: the same characteristics at each level have different metrics or observation value ranges, and the tools at each level also distort the results of the observations.
\subsubsection{Why WPC is essential?}~\label{Motivation_WPC}

Simply performing single-level workload characterization alone will lead to misleading or contradicted conclusions. On the other hand, integrating three-level profiling data and performing the statistics-based analysis (such as the Pearson correlation analysis) relies on the observation samples and can only obtain the statistics characteristics across stacks. Our case study in Section~\ref{Motivation_integration} confirms that the simple integration of three-level profiling data does not help gain new insights. Up to now, we can not answer our proposed workload characterization issue: can we uncover each critical component across the stacks contributing what percentages to any specific bottleneck?
So, We propose the WPC methodology and build the tool to tackle the above challenge.
\section{The WPC Methodology and Tool}~\label{TheMedthodSection}

\begin{figure}[h]
\centering
\includegraphics[scale=0.65]{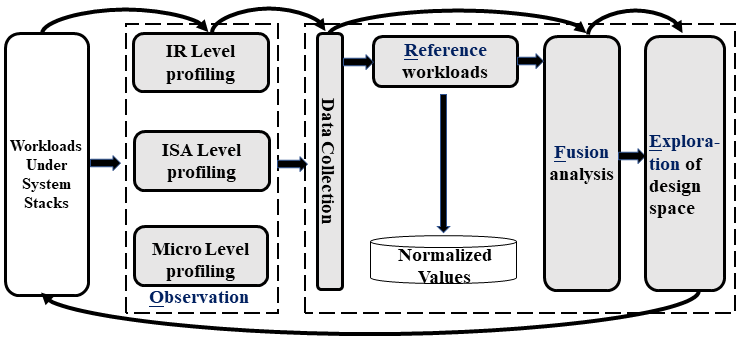}
\caption{The WPC methodology}
\label{method1fig0001meth}
\end{figure}

Fig.~\ref{method1fig0001meth} shows the WPC methodology framework. The WPC methodology is an iterative ORFE loop consisting of four steps: \underline{O}bservation, \underline{R}eference, \underline{F}usion, and \underline{E}xploration. First, we observe workload characteristics at three primary levels: IR, ISA, and microarchitecture levels, and integrate the observation data. Second, we propose the standard reference workloads for quantifying the variations of the complex workloads' characteristics across system stacks. These standard reference workloads are analogies with the reference machine used in SPEC CPU~\cite{spec}, which forms the yardstick to evaluate a target machine. Third, the fusion analysis fuses and normalizes different levels of profile data to depict workload characteristics in a combined and comprehensive way. Fourth, we explore the software and hardware co-design space and further investigate the impacts of critical components.

\subsection{The WPC methodology}

\subsubsection{Observation}~\label{Method_WPC1}

In WPC, we observe workload characteristics at three primary levels: IR, ISA, and microarchitecture levels. The IR level analysis targets the programming framework and language beyond the runtime environments. In contrast, the ISA level analysis consists of the runtime environments and the ISAs, beyond the OS. The microarchitecture level analysis is microarchitecture dependent and affected by the OS. Furthermore, we can analyze more fine-grained characteristics at the observation level to obtain more accurate characteristics. For example, we can decouple the programming framework and language by analyzing the IR level profiling data.

Within each level, we observe the workloads from the perspectives of instruction locality (Frontend related metric), data locality (Backend related metric), and branch locality (Speculation related metric), each of which is an essential metric that identifies the critical pipeline bottleneck.

We choose instruction Reuse\_Dist~\cite{1628963}, instruction Reuse\_Dist, and L1I Cache MPKI as metrics to measure the instruction locality at the IR, ISA, and microarchitecture levels, respectively. The details are in Section.~\ref{MotivationMetrics}.
At the IR, ISA, and microarchitecture levels, we choose the data Reuse\_Dist~\cite{1628963}, the data Reuse\_Dist, and the L1D Cache MPKI, respectively, as the data locality metrics. Among them, the data Reuse\_Dist is the average distance between two consecutive accesses to the same data address. A larger data Reuse\_Dist value implies a lower data locality. The data Reuse\_Dist is calculated at the IR and ISA levels. We use Perf to measure the L1D Cache MPKI at the microarchitecture level. We adopt branch locality to measure the control category. Branch entropy~\cite{10.5555/1287050.1287068}, Branch entropy, and branch MPKI  measure the branch locality at the IR, ISA, and microarchitecture characteristics.
Branch entropy is the linear branch entropy, calculated as
$H(X)=2*min(p(x),1-p(x))$, where $p(x)$ is the taken probability of the branch $x$. We use Perf to measure the branch MPKI at the microarchitecture level.

\subsubsection{Reference}~\label{Method_WPC2}

At the observation step, the IR-level profiling omits the characteristics of the third-party libraries, such as the runtime libraries and OS libraries. The ISA-level analysis cannot capture the OS systems function calls because all existing ISA-level profiling tools run as a process on the OS. The microarchitecture-level analysis is performed on a specific microarchitecture. To avoid distortion from the observation tools and quantify the characteristics variations across the system stacks, we
propose a series of standard reference workloads with three fundamental properties. First, their behaviors are deterministic and consistent. Second, the slight deviation of behavior across different stacks is explainable. Third, they are portable and deployed on any platform. Then we can use them to quantify the characteristics variations across the system stacks and the characteristics variations of the complex workloads.

We implement three standard reference workloads: the standard data locality reference workload, the standard instruction locality reference workload, and the standard branch locality reference workload. Table~\ref{table-Workloads-SRW} shows the overview of the three workloads. All workloads avoid using third-party libraries and  OS-intensive function calls. All workloads' parameters are configurable to cover a wide characteristics space, from an excellent locality to a poor locality.

The implementation details of the standard reference workloads are as follows:

\textbf{The standard data locality reference workload.} We implement the standard data locality workload by controlling the data Reuse\_Dist of the workload. The data Reuse\_Dist is the number of memory access instructions in the interval between two consecutive accesses to the same data address. Inspired by~\cite{McCalpin2007},  which allocates a large data array and performs random access to the array. Controlling Reuse\_Dist of the workloads is achieved by controlling the access size of the data array. We use $X$ to refer to the access size of the data array. To keep it concise, in the rest of this paper, $X$ refers to the controlled parameter. Please note that $X$ has different meanings in different standard reference workloads. The data Reuse\_Dist expectation is calculated as  $\lim_{n \to \infty} \frac{1}{X} \sum_{i=1}^{n} i (\frac{X-1}{X})^{i-1} = X$, where $n$ refers to the access number. The theoretical Reuse\_Dist range of the standard data locality reference workload is 1 to 2,000,000. As a value of Reuse\_Dist, 2,000,000 is large enough.

\textbf{The standard instruction locality reference workload.} We implement the standard instruction locality reference workload by controlling the instruction Reuse\_Dist of the workload. The instruction Reuse\_Dist is the number of instructions in the interval between two consecutive accesses to the same instruction address. Inspired by~\cite{ayers2019asmdb}, we define lots of void functions, and a large array is allocated to save those function entry addresses. Then we randomly access this array and call the function. Similar to the standard data locality reference workload, we tune the value of Reuse\_Dist by controlling the access size of the array. Here, $X$ refers to the access size of the array. Since each function is translated into multiple instructions, the expectation of the instruction Reuse\_Dist is $X*b$, where $b$ is the number of translated instructions (in our LLVM-based implementation, the $b$ is 5). The theoretical Reuse\_Dist range of the standard instruction locality reference workload varies from 5 to 2,000,000. As a value of Reuse\_Dist, 2,000,000 is large enough.

\textbf{The standard branch locality reference workload} A random number is sampled from the range of $[0,m]$,  and then compared with the threshold $X$, which also ranges from 0 to m, to determine whether the branch will jump. The branch will jump if it is less than the threshold $X$. Otherwise, the branch will not jump. In our implementation, we fix the $m$ (the $m$ is set to 1000) and change the threshold value to control the branch jump rate. The probability of the branch jump is calculated as $X/m$. We use linear branch entropy as the metric of branch locality, which is defined as $2*min(X/m,1-X/m)$. The theoretical linear branch entropy range of the standard branch locality reference workload varies from 0.002 to 1, and 1 is the upper bound of the linear branch entropy.


\begin{table}[ht]
\small
\centering
    \caption{The overview of the standard reference workloads. To keep concise, we use $X$ to refer to the controlled parameter. $X$ has a different meaning in different standard reference workloads. $b$ and $m$ are constants. }
    \label{table-Workloads-SRW}
    \centering
    \begin{tabular}{|c|c|c|}
        \hline
     \textbf{Standard reference workload} &\textbf{Metric}  & \textbf{Theoretical prediction}  \\
     \hline
    The data locality workload & Reuse Distance  & $X$\\\hline
    The instruction locality workload. & Reuse Distance   & $X*b$ \\\hline
    The branch locality workload & Linear Entropy  & $2*min(X/m,1-X/m)$\\
     \hline
    \end{tabular}
\end{table}

\subsubsection{Fusion}~\label{Method_WPC3}

We propose two fusion analysis methods: statistics-based analysis and the standard reference workloads-based analysis.

In the statistics-based analysis, we use statistical analysis to understand the similarities/variations of the workload characteristics across the stacks, such as the correlation analysis in Section~\ref{Motivation_integration}.

The statistics-based analysis can identify the correlation factors across stacks but can not quantify the contributing factors of the components across stacks. So we propose the standard reference workloads-based analysis. In the standard reference workloads-based analysis, we choose the standard reference workloads as the baseline for normalization. We calculate the relative values of the analyzed workload against the standard reference workloads. Then we can quantitatively analyze the characteristic variations among these levels. We propose the normalized impact factors to reflect the effects of different components across the system stacks on the characteristics of the workload.

The normalized impact factor is calculated according to Equation~\ref{formula-1}. $X_i$ is the characteristic value of the analyzed workload at the IR, ISA, or microarchitecture levels. $S_i$ is the corresponding value of the standard reference workload; $R_i$ is the relative value at the IR, ISA, or microarchitecture levels, calculated as the ratio of the characteristic value of the analyzed workload to the corresponding value of the standard reference workload at each level. $I_i$ is the normalized impact factor. In Equation~\ref{formula-1}, $n$ is 3 (three observation levels).

    \begin{equation}
    \label{formula-1}
     I_i= R_i/ \sum_{i=1}^{n} R_i\;, \quad R_i=X_i/S_i
    \end{equation}

For example, the observation values of instruction locality of the Bayes-Hadoop workload at the IR, ISA, and microarchitecture levels are 49086 (Reuse\_Dist), 8824 (Reuse\_Dist), and 16.9 (MPKI), respectively. The instruction locality of the standard instruction locality reference workload is 2040 (Reuse\_Dist), 2421 (Reuse\_Dist), and 0.43 (MPKI), respectively. The relative values are 24.1, 3.6, and 39.4. Then the normalized impact factor of the IR-dependent category, ISA-dependent category, and microarchitecture-dependent category
is 0.36, 0.05, and 0.59, respectively. More details are shown in Section~\ref{EvaluationsScaleOut}.

\subsubsection{Exploration}~\label{Method_WPC4}

The fusion analysis results provide feedback for exploring the software and hardware design space. For example, the fusion analysis reveals that the Java language and JVM incur higher L1I Cache Misses of the scale-out Hadoop workloads at the microarchitecture level. So, we change the programming framework from MapReduce to MPI, another scale-out framework, and the language from Java to C. The L1I Cache MPKI of the Bayes-MPI workload eliminates by 77\%. More details can be found in Section~\ref{EvaluationsScaleOut}.

\subsection{The WPC tool}~\label{HierarchicalTools}
Besides the standard reference workloads, the other primary WPC modules include a multi-level profiler and a performance data analyzer. The multi-level profiler integrates LLVM, Hotspot, Pin, DynamoRIO, and Perf, profiles the workloads, and gathers performance data. The performance data analyzer will collect all the data generated by the multi-level profiler and store them in the database. The performance data analyzer reads raw data from the database and analyzes the multi-level performance data. Its figure plotter plots different figures, facilitating users to analyze the performance metrics. For example, the performance data analyzer can automatically perform standard reference workloads-based analysis described in Section~\ref{Method_WPC3}.

The primary overhead of WPC is at the observation step. In our experiments, analyzing the instruction locality of the Bayes-Hadoop workload consumed 290, 1069, and 13 seconds at the IR, ISA, and microarchitecture levels.  Accordingly,  using state-of-practice tools, the time is 1069 seconds using Intel Pin for ISA-level profiling  and 13 seconds using Intel VTune for microarchitecture-level profiling. WPC supports performing three-level observations in parallel. So the total time consumed by WPC is around 1069 seconds, which is acceptable.

\section{Evaluations}\label{TheEvaluationsSection}

The goal of the evaluations is four-fold. First, we validate the standard reference workloads and guide how to choose the proper X value. Second, we evaluate two ISA platforms, the Intel Xeon and AArch64 platforms, with the standard reference workloads. Third, we characterize the scale-out Hadoop workloads by comparing them with the SPECrate workloads. Fourth, we discuss and emphasize the advantages of WPC to quantify the impacts of the language, programming framework, and runtime environment.


\subsection{Experimental configurations}\label{EvaluationsConfigurations}

In most experiments, we deploy the workloads on the Intel Xeon Gold 5120T processor equipped with 384 GB memory and an 8 TB disk. The OS is Ubuntu 16.04. WPC integrates LLVM (version 9.0), Hotspot (version 1.8), DynamoRIO (Version 9.0.1), Pin (Version 3.17), and Perf (Version 4.4) as the profile tools. We repeat each experiment more than three times and report the average values. We also evaluate two ISAs -- typical RISC (AArch64) and CISC (X86-64) ISAs. For an AArch64 microarchitecture, we evaluate the AArch64 Kunpeng 920 processor, equipped with 64 GB memory and 8 TB disk. For evaluating different microarchitectures, we also evaluate the Intel Xeon E5645 processor, equipped with 64 GB memory and 8 TB disk. In the default settings,  we turn off the hardware prefetcher.

We evaluate the total 23 throughput-oriented SPECrate workloads from SPEC CPU2017 and run the official applications with the reference input. We choose eight typical scale-out workloads implemented with Hadoop frameworks. Bayes-Hadoop takes wiki news ~\cite{Wiki20News} as the input. PageRank-Hadoop analyzes PEGASUS data~\cite{5360248}. Kmeans-Hadoop takes the graph vertex data~\cite{bigdatabench} as the input. CF-Hadoop performs collaborative filtering analysis on the MovieLens data~\cite{10.1145/2827872}. Sort-Hadoop, Grep-Hadoop, MD5-Hadoop~\cite{rachmawati2018comparative}, and CC-Hadoop~\cite{kolb2014iterative} are primary operations in the big data processing.

\subsection{Validation of the standard reference workloads}~\label{EvaluationsMO}

\subsubsection{Validating  the actual values against the theoretical values at the IR level}~\label{EvaluationsMO-1}

\begin{figure}[h]
\centering
\includegraphics[scale=0.4]{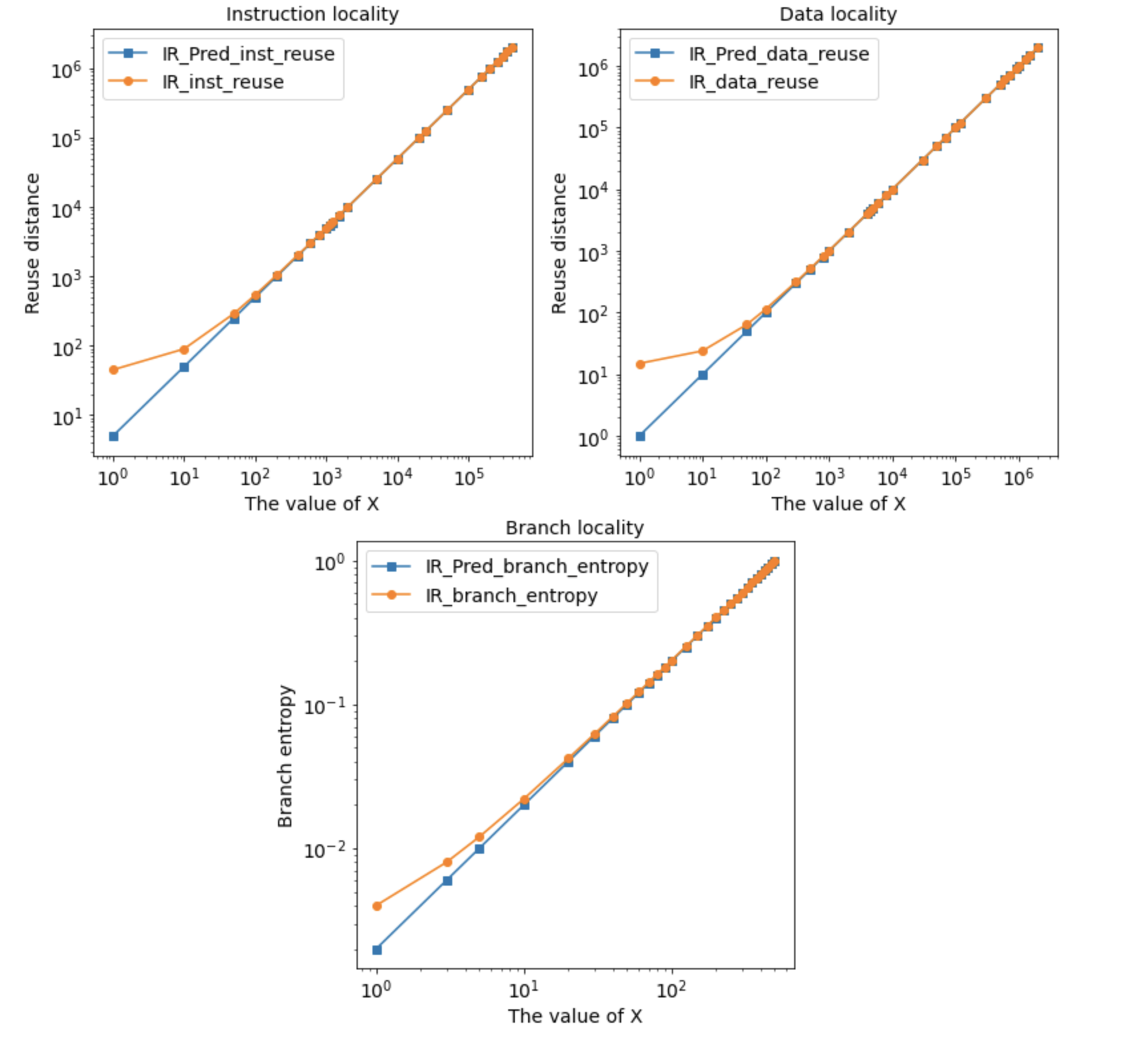}
\caption{The theoretically predicted values vs. the actual values at the IR level when tuning the parameters $X$ in the three standard reference workloads.}
\label{ValMOIRSorceCode}
\end{figure}

Fig.~\ref{ValMOIRSorceCode} compares the theoretically predicted values (according to Table~\ref{table-Workloads-SRW}) against the actual values at the IR level when tuning the parameters X in the standard reference workloads. We find that they contain almost overlapped curves and reflect consistent characteristics. Against the theoretically predicted values, the average error~\cite{RelativeError} of the actual instruction locality of the standard reference workload is 6.7\% when X ranges from 1 to 400,000. Not including the range where X is less than 400 (more significant deviation), the average error is below 0.5\%. Analogously, against the theoretically predicted values, the average error of the actual data locality is 6.9\% when X ranges from 1 to 2,000,000. Not including the range where X is less than 800 (more significant deviation), the average error is below 0.3\%. Against the theoretically predicted values, the average error of the actual branch locality is 4.3\% when X ranges from 1 to 500. Not including the range where X is less than 60 (more significant deviation), and the average error is below 0.6\%. Fig.~\ref{ValMOIRSorceCode} indicates that the design and implementation of the standard reference workload meet the design targets.

\subsubsection{The consistency across the IR, ISA, and microarchitecture levels} ~\label{EvaluationsMO-2}

\begin{figure}[h]
\centering
\includegraphics[scale=0.65]{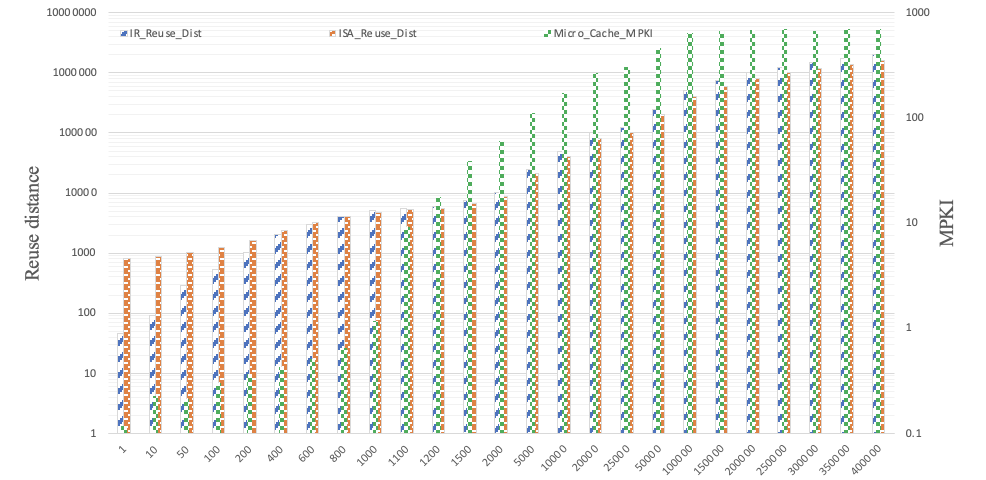}
\caption{The instruction localities of the standard instruction locality reference workload at three levels.}
\label{ValMOInsLo}
\end{figure}

\begin{figure}[h]
\centering
\includegraphics[scale=0.7]{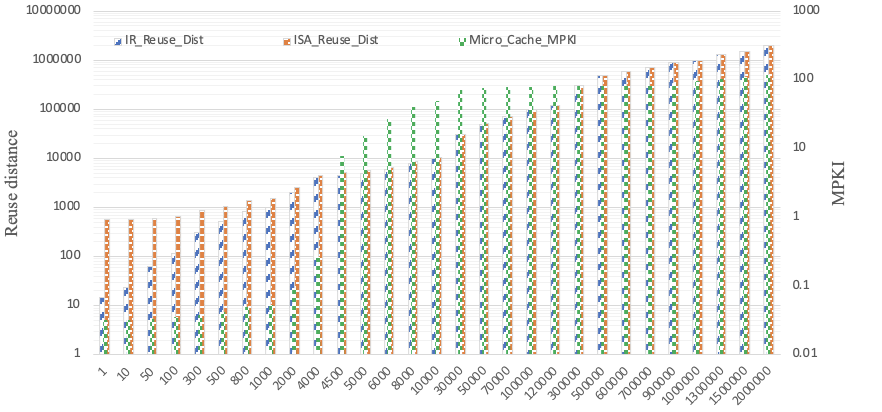}
\caption{The data localities of the standard data locality reference workload at three levels.}
\label{ValMODataLo}
\end{figure}

\begin{figure}[h]
\centering
\includegraphics[scale=0.7]{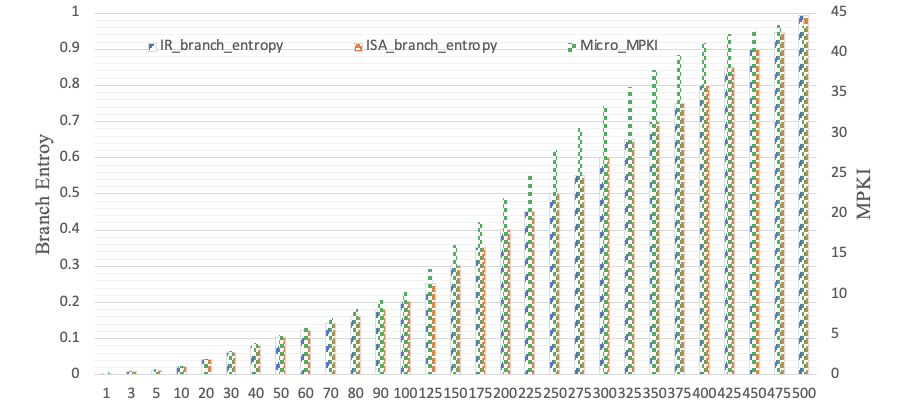}
\caption{The branch localities of the standard branch locality reference workload at three levels.}
\label{ValMOBranchLo}
\end{figure}

Fig.~\ref{ValMOInsLo}, Fig.~\ref{ValMODataLo}, and Fig.~\ref{ValMOBranchLo} show instruction, data, and branch localities at three levels, respectively. The microarchitecture platform is Intel Xeon Gold 5120T. First, the instruction, data, and branch localities keep consistent among the three levels. Second, the correlation coefficients between the branch localities at the IR and ISA levels and between them at the ISA  and microarchitecture levels are all 0.99. Third, the correlation coefficients between the instruction localities and data localities at the IR and ISA levels are all 0.99. The correlation coefficient between the instruction localities at the ISA and microarchitecture levels is 0.87. The correlation coefficient between the data localities at the ISA and microarchitecture levels is 0.74, which is lower because the cache size of L1I Cache and L1D Cache affect the microarchitecture performance. When $X$
equals 1000, it indicates the first working set for the standard instruction locality reference workload, corresponding with the L1I Cache size (32KB), as the designed instruction Reuse\_Dist value is 5X. When $X$ equals 4000, it indicates the first working set for the standard data locality reference workload, which corresponds with the L1D Cache size (32KB), as the designed data Reuse\_Dist value is X. Above all, the standard reference workloads remain consistent among the IR, ISA, and microarchitecture levels. Meanwhile, they are deterministic and explainable.

\subsubsection{Guidance on the selection of X value}
~\label{GuidanceXvalues}

According to our experiment requirements, we define two criteria to select a proper $X$ number: (1) The average error between the actual IR number and the theoretical prediction is less than 2\%. (2) The actual IR number should achieve the best locality within the range.
Fig.~\ref{ValMOInsLo} shows the changing trend of the instruction locality under different X values. According to criterion (1), 400 is the first value with an average error below 2\%. Thus we should choose the X value greater or equal to 400. According to criterion (2), we set $X$ as 400 in the standard instruction locality reference workload considering these constraints. Analogously, we choose the proper X value for the standard data locality reference workload and the standard branch locality reference workload, which are  800 and 60, respectively.

\subsection{Evaluating different ISAs with the standard reference workloads}~\label{EvaluationsISA}

We use the standard reference workloads to measure the variations of workload characteristics across different ISAs -- typical RISC (AArch64) and CISC (X86-64) ISAs. The microarchitecture platforms are Kunpeng 920 for  AArch64 and Intel Xeon Gold 5120T for X86-64.
WPC uses DynamoRIO as the ISA-level profiling tool  and Perf as the microarchitecture-level tool. We do not report the value of branch locality because this feature of DynamoRIO is under development.

\subsubsection{Data locality} ~\label{EvaluationsMO-DataLocality}

\begin{figure}[h]
\centering
\includegraphics[scale=0.6]{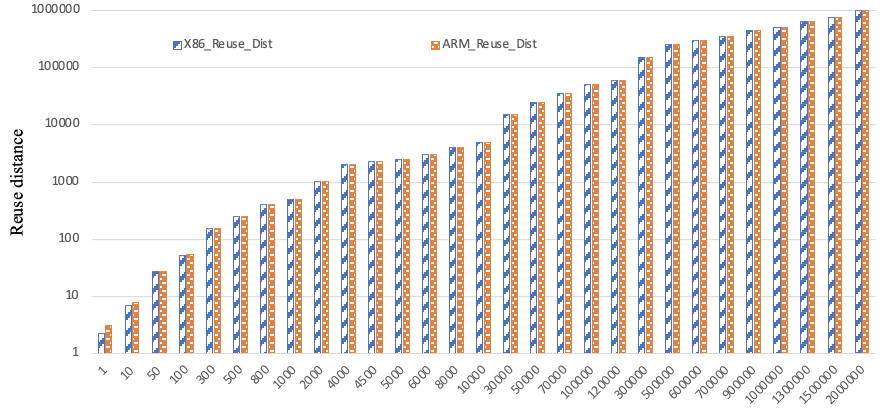}
\caption{The data localities of the standard reference workloads at the AArch64 and X86-64 ISAs.}
\label{MoDiffISAData}
\end{figure}

The numbers of the standard data locality reference workload remain consistent between AArch64 and X86-64 at the ISA level in Fig.~\ref{MoDiffISAData} and have similar values. It implies that different ISAs have little impact on data locality.

We analyze the microarchitecture characteristics. When $X$ equals to 8000, there is the first working set for the standard data locality reference workload on AArch64. Contrasted, when $X$ equals 4000, there is the first working set for the standard data locality reference workload on X86-64. The former is twice as much as the latter because  Kunpeng 920 (the AArch64 platform) has  a 64KB L1D Cache, twice as much as the 32KB L1D Cache size of Gold 5120T (the X86-64 platform).
Finally, the correlation coefficients between the data localities at the IR and X86-64 levels and between the data localities at the IR and AArch64 levels are all 0.99, implying the standard data locality reference workload is not affected by ISAs significantly.


\subsubsection{Instruction locality}~\label{EvaluationsMO-InsLocality}

\begin{figure}[h]
\centering
\includegraphics[scale=0.8]{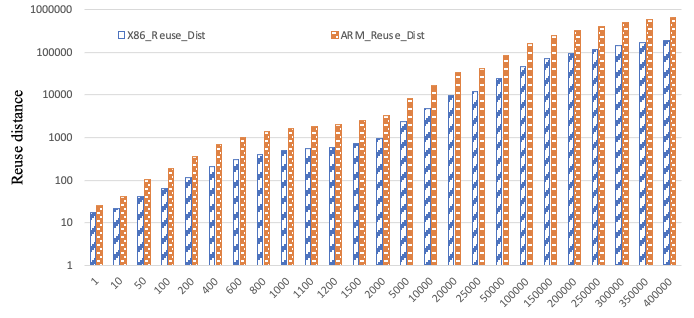}
\caption{The instruction localities of the standard reference workloads at the AArch64 and X86-64 ISAs}
\label{MoDiffISAIns}
\end{figure}

For the instruction locality, there is a similar tendency but different numbers between AArch64 and X86-64 at the ISA level, as illustrated in Fig.~\ref{MoDiffISAIns}. We find that ISAs significantly impact the instruction locality, and the average gap is 3.4 times between AArch64 and X86-64. The RISC ISA of AArch64 leads to more instructions and a worse locality. In our evaluation with the standard instruction locality reference workload, the instructions are 1E10 on average for AArch64 and 6.8E9 on average for X86-64, with a gap of 1.5 times. More instructions would lead to a worse instruction locality for the same workload implemented with different machine codes.

We analyze the microarchitecture characteristics. When  $X$ equals 2000,  there is the first working set for the standard instruction locality reference workload in AArch64. Contrasted, when $X$ equals 1000, there is the first working set for the standard instruction locality reference workload in X86-64. The former is twice as much as the latter because   Kunpeng 920 (the AArch64 platform) has  a 64KB  L1I Cache, twice as much as the 32KB L1I Cache size of Gold 5120T (the X86-64 platform).
Finally, the correlation coefficients between the instruction locality at the IR and the X86-64 levels and between the instruction locality at the IR and the AArch64 are all 0.99, implying that the standard instruction locality reference workload is not impacted by ISAs significantly.

\subsection{Analyzing the scale-out Hadoop workloads and the SPECrate workloads }~\label{EvaluationsFusion}

We perform workload analysis on Intel Xeon Gold 5120T. We also report the microarchitecture level data on  Intel Xeon E5645. We do not report the data on the ARM platform because the ISA-level tool, DynamoRIO,  does not support the scale-out Hadoop workloads on the ARM platform very well, and DynamoRIO has inferior performance when running Java~\cite{farhaditrace}. Also, we find it hard to run the scale-out Hadoop workloads when instrumented with  DynamoRIO on our ARM platform.


\subsubsection{Instruction Locality}~\label{EvaluationsMO441-InsLocality}

Fig.~\ref{motivationfig0001} reports the instruction localities across the IR, ISA, and microarchitecture levels. From Fig.~\ref{motivationfig0001}, we can find that the scale-out Hadoop workloads and SPECrate workloads have similar values at the IR level but different values at the microarchitecture level. The scale-out Hadoop workloads have high L1I Cache Misses.

\begin{figure}[h]
\centering
\includegraphics[scale=0.25]{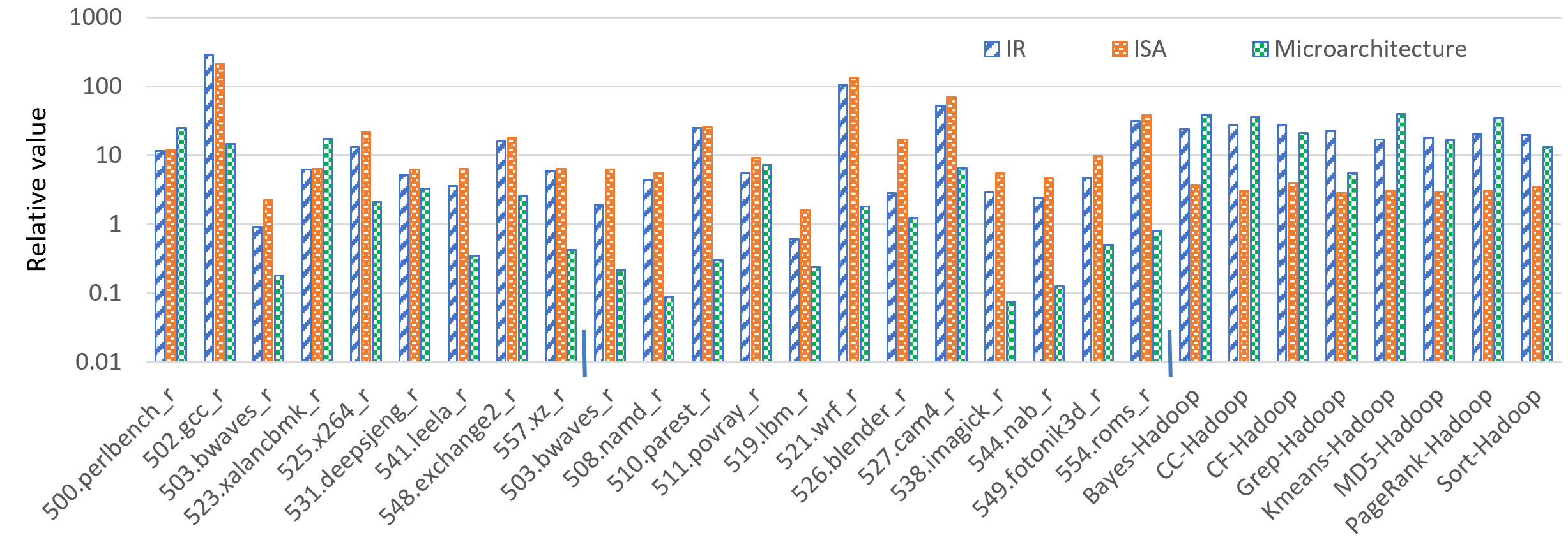}
\caption{The relative instruction localities of the target workloads against the standard reference workload}
\label{EvaluationFusion0002}
\end{figure}

Furthermore, We calculate the relative values of the target workload against the standard instruction locality reference workload. We choose X as 400 for the standard instruction locality reference workload, described in Section~\ref{GuidanceXvalues}. The instruction locality of the standard instruction locality reference workload at the IR, ISA, and microarchitecture levels is 2040 (Reuse\_Dist), 2421 (Reuse\_Dist), and 0.4 (MPKI), respectively.
Fig.~\ref{EvaluationFusion0002} shows the relative value. We can quantify workload characteristics variations across the system stacks. For example, for the 500.perlbench\_r, the variation from the IR level to the ISA level is 1.01 times, while that from the ISA level to the microarchitecture level is 2.1 times. From Fig.~\ref{EvaluationFusion0002},
the SPECrate workloads' observation values keep consistent from the IR level to the ISA level (the average gap is 1.5 times). From the ISA level to the microarchitecture level, the observed locality value improves (the average gap is 0.34 times), indicating that the instruction cache mechanisms of microarchitecture are effective. On the other hand, the observation values of the scale-out Hadoop workloads have significant fluctuation among different levels. The observation values significantly decrease from the IR level to the ISA level (the average gap is 0.15 times) and increase dramatically from the ISA level to the microarchitecture level  (the average gap is 7.9 times). As we observe that the ISA level is a critical one in significant fluctuation, we speculate that the ISA-dependent components of the scale-out Hadoop workloads are the primary ones that significantly contribute to the fluctuation across levels.

As the ISA-dependent component, JVM (java runtime environment) includes many mechanisms to manage the Java virtual machine for the Java program execution. For example, the JIT mechanism implements instructions' binary translation and execution in time, closely related to the instruction locality. We turn the JIT off and
find that the SPECrate workloads and the scale-out Hadoop workloads  have similar L1I Cache MPKI, which is 4.1 and 4.4, respectively.


\subsubsection{Data Locality}~\label{DataLocalSubsub}

We measure data locality across the IR, ISA, and microarchitecture levels. We do not report the value of the scale-out Hadoop workloads at the IR level because the corresponding profiling tool is under development. We find that the scale-out Hadoop and the SPECrate workloads have different values at the ISA level (the gap is one magnitude of order) but have similar values at the microarchitecture level. For the SPECrate workloads, the correlation coefficient between the data localities at the IR level and ISA levels is 0.88, which implies a strong positive correlation. But the correlation coefficient between the data localities at the ISA and microarchitecture (Intel Xeon Gold 5120T) levels is 0.2, and the value is 0.14 on Intel E5645. At the same time, the data locality of the scale-out Hadoop workloads has no significant similarity between the ISA  and microarchitecture levels, and the correlation coefficient of the data locality is only 0.03 ( X86-64 vs. Intel Gold 5120T) and 0.05 (X86-64 vs. Intel E5645).

\begin{figure}[h]
\centering
\includegraphics[scale=0.28]{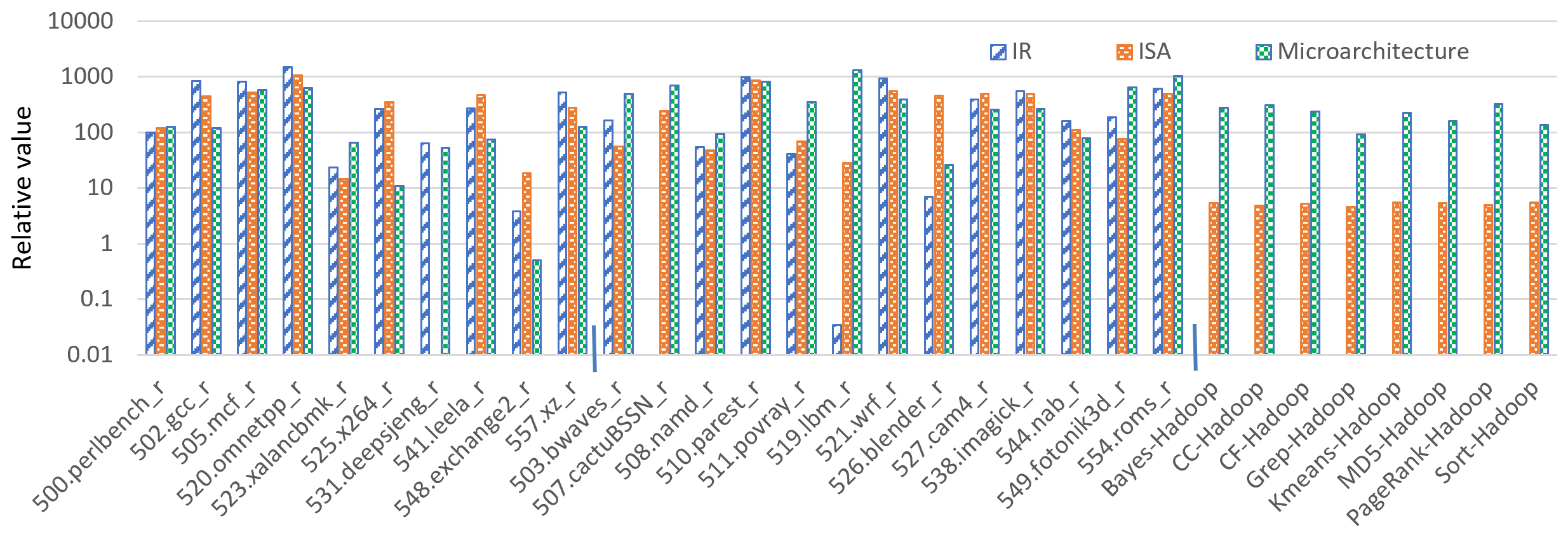}
\caption{The relative data localities of the target workloads against the standard reference workload}
\label{EvaluationFusion0005}
\end{figure}

Furthermore, We calculate the relative values of the target workload against the standard data locality reference workload. We choose the X as 800 for the standard reference workload, described in Section~\ref{GuidanceXvalues}. The data locality of the standard reference workload is 814 (Reuse\_Dist), 1367 (Reuse\_Dist), and 0.05 (MPKI), respectively. From Fig.~\ref{EvaluationFusion0005}, we find that the SPECrate workloads observation values keep consistent between the IR level and ISA level except for the 519.lbm\_r and 526.blender\_r workloads, and their average gap is 1.1 times. On the other hand, the data locality of the scale-out Hadoop workloads have
less fluctuation at the ISA level (the coefficient of variation~\cite{Brown1998} is only 0.05).

\begin{figure}[h]
\centering
\includegraphics[scale=0.3]{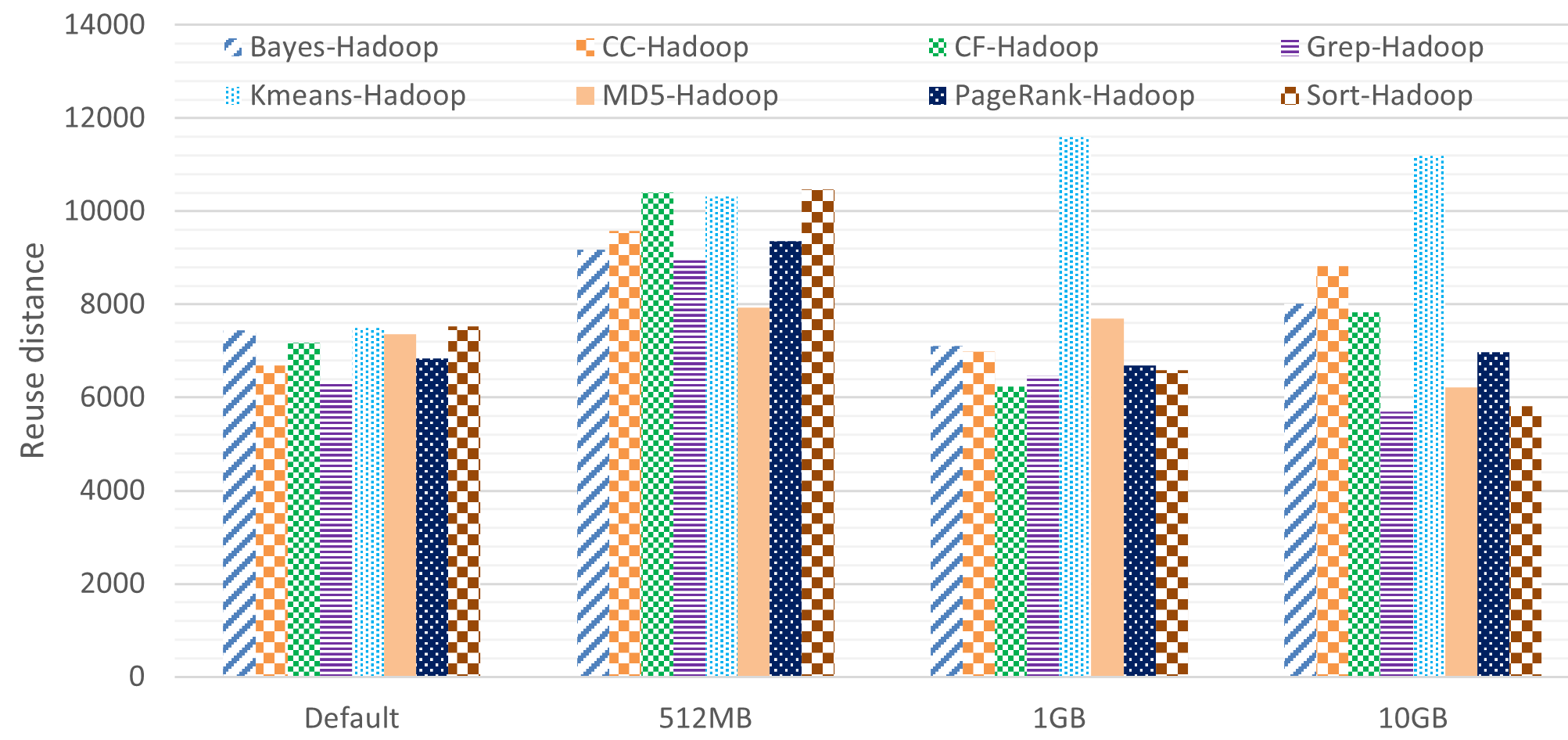}
\caption{The data locality values of the Hadoop workloads at the ISA level under different JVM heap sizes}
\label{EvaluationFusion0006}
\end{figure}

To reveal the reason for the less data locality fluctuation of the scale-out Hadoop workloads at the ISA level, we analyze the impact of the maximum heap size parameter setting -- a standard setting of JVM for memory management -- on the Java workloads. Fig.~\ref{EvaluationFusion0006} presents the results. On the one hand, different heap settings significantly impact data locality at the ISA level. On the other hand, the default heap size settings can achieve the best results and keep less fluctuation. Hence, JVM's automatic heap management mechanism allows Java workloads to obtain better data locality at the ISA level, reducing data locality value fluctuation.

\subsubsection{Branch Locality}~\label{EvaluationsMO443-BranchLocality}

The branch locality of the SPECrate INT workloads has similarities across the stacks. The correlation coefficients between the branch localities at the IR and ISA levels,  at the ISA and microarchitecture (Intel Gold 5120T architecture) levels, and at the ISA  and microarchitecture (the Intel E5645 architecture) levels are 0.99, 0.71, 0.85, respectively. The branch locality of the SPECrate FP workloads has similarities between the IR and ISA levels, and the correlation coefficient is 0.85.
At the same time, the branch locality of the scale-out Hadoop workloads has no  similarity between the IR  and ISA levels, and the correlation coefficient is -0.003.

\begin{figure}[h]
\centering
\includegraphics[scale=0.28]{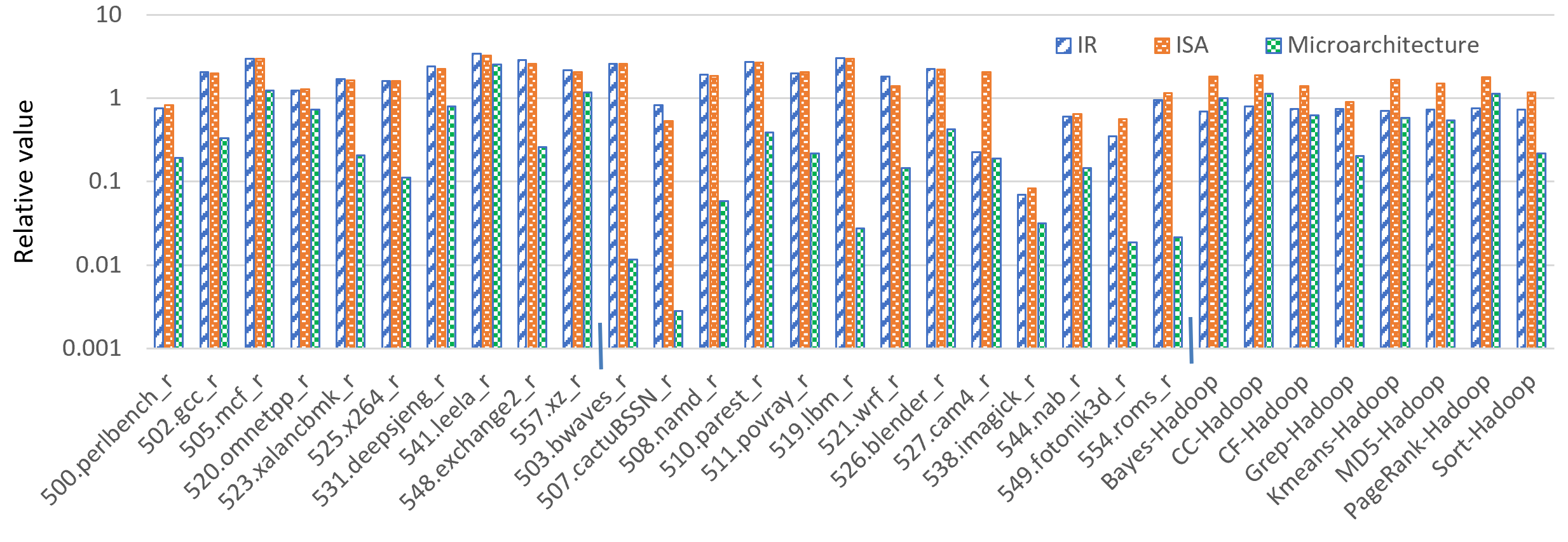}
\caption{The relative branch localities of the target workloads against the standard reference workload}
\label{EvaluationFusion0009}
\end{figure}

We calculate the relative values of the target workload against the standard branch locality reference workload. We choose  X as 60 for the standard reference workload, described in Section~\ref{GuidanceXvalues}. The branch locality of the standard  reference workload is 0.12 (branch entropy), 0.12 (branch entropy), and 6.0 (MPKI), respectively. From Fig.~\ref{EvaluationFusion0009}, we can see that the SPECrate workloads observation values keep consistent from the IR to the ISA levels except for the 527.cam4\_r workload~\footnote{the 527.cam4\_r  workload mainly call the cam4 packet, while the IR level analysis omits the characteristics of the third-party libraries}, and their average gap is 1.01 times. Furthermore, the locality improves from the ISA  to the microarchitecture levels, indicating the branch prediction mechanisms of microarchitecture are effective. On the other hand, the branch locality of the scale-out Hadoop workloads have
less fluctuation at the IR level, and the coefficient of variation is only 0.04.

The main reason for the fewer branch locality fluctuations at the IR level is that there are many JVM-related codes (the management codes for the JVM) in the Java workload. The code of the Java workload can be divided into JVM-related code and workload-specific code. By analyzing the bytecode traces at the IR level, we find that the standard deviation of the branch entropy of the entire bytecodes (plus workload-specific with JVM-related)  is only 0.03. In contrast, the standard deviation of the branch entropy of the workload-specific bytecodes is 0.06, twice the former. Bytecodes of the JVM-related codes have similar branch-jumping behaviors and thus make the scale-out Hadoop workloads' IR level branch entropy more similar.

\subsection{Exploring the optimizations}
We investigate the impacts of the single-level and across-stack optimizations as examples to demonstrate the exploration. The microarchitecture platform is Intel Xeon Gold 5120T.

\subsubsection{The single level optimization}
~\label{EvaluationsMO451-1}

The hardware prefetcher is a default optimization for microarchitecture. For the standard instruction locality reference workload, when the prefetcher is on or off,  L1I Cache MPKI differs by an average of 2.7\%. For the standard data locality reference workload, when the prefetcher is on or off, the average difference of L1D Cache MPKI is 8.5\%. Furthermore, for the SPECrate workloads,  when the prefetcher is on or off,  L1I Cache MPKI differs by an average of 13\%;  the average difference between  L1D Cache MPKI is 11\%.
For the scale-out Hadoop workloads,  when the prefetcher is on or off, L1I Cache MPKI differs by an average of 1\%;  the average difference between  L1D Cache MPKI is 8\%.


\subsubsection{Optimization across the stacks }~\label{EvaluationsCrossStackTune}

\begin{figure}[h]
\centering
\includegraphics[scale=0.45]{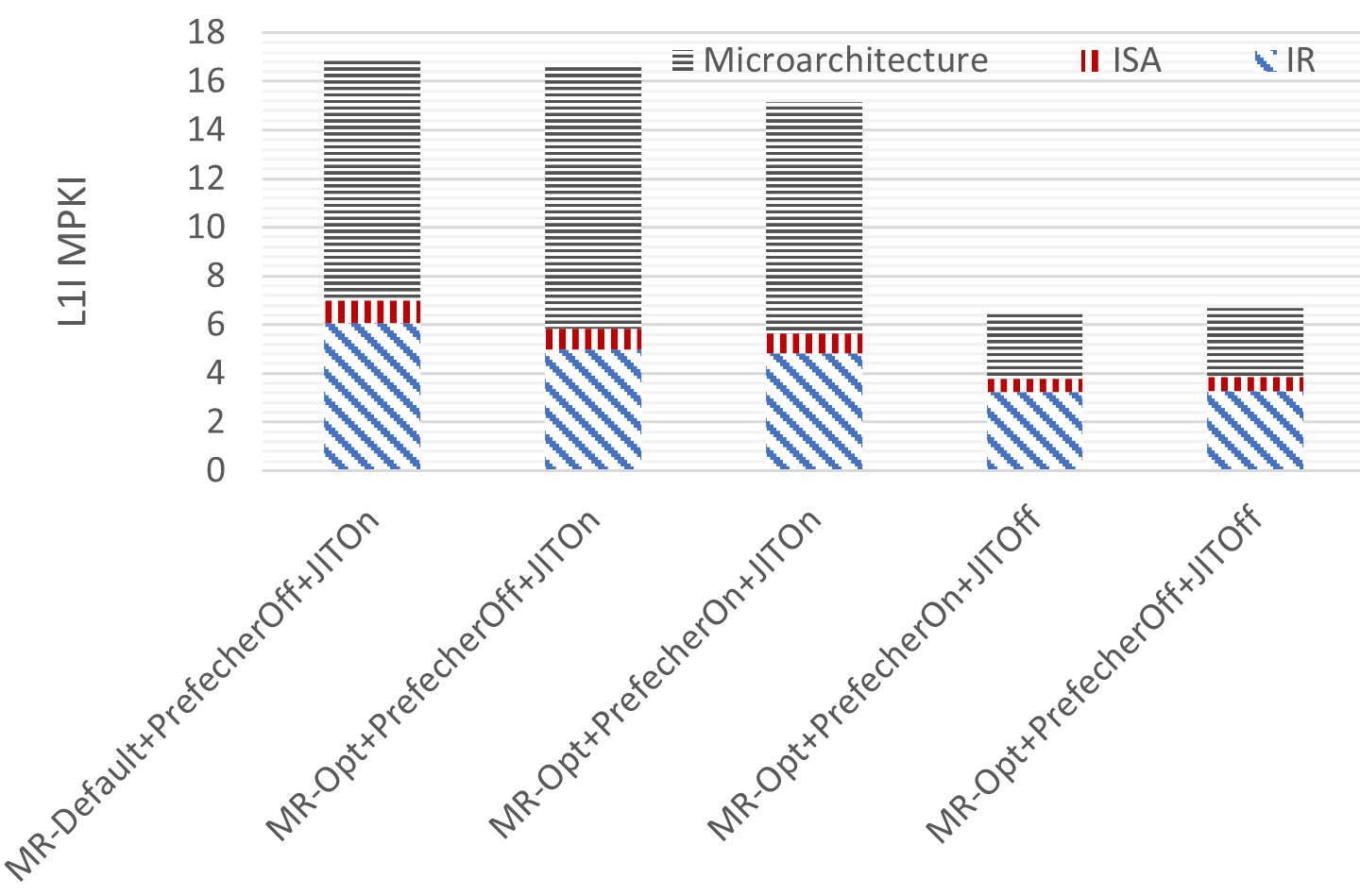}
\caption{Break down the normalized L1I Cache Misses under different configurations.}
\label{EvaluationFusion0010}
\end{figure}

There are many tuning parameters at different levels across the system stacks. We evaluate the effects of tuning across stacks with different configurations, turning on or off the prefetcher at the microarchitecture level, turning on or off JIT at the ISA level, and changing the MapReduce setting at the IR level -- the start time of the reduce tasks and the number of tasks. We calculate the normalized impact factors of each component in the configuration according to Equation~\ref{formula-1}. For clarity, Fig.~\ref{EvaluationFusion0010} reports the normalized L1I Cache MPKI, which is the normalized impact factors multiplied by the L1I Cache MPKI of each configuration. From Fig.~\ref{EvaluationFusion0010}, different configurations produce different optimization efforts. In Configurations One and Two, comparing the MapReduce settings (Column I: the default MapReduce settings; Column II: MapReduce Optimization settings), the IR-level and ISA-level values --normalized  L1I Cache MPKI -- improve (Column II is better than Column I), and the microarchitecture level value keeps consistent. In Configurations two and Three, comparing the MapReduce Optimization settings with the prefetcher on or off (Column II: MapReduce Optimization settings with the prefetcher off; Column III: MapReduce Optimization settings with the prefetcher on), the IR level and the ISA-level value keep consistent.
In contrast, the microarchitecture-level value improves  (Column III is better than Column II). In Configurations three and Four, comparing the MapReduce Optimization setting with JIT on or off and the prefetcher on (Column III: the MapReduce Optimization setting with JIT on and the prefetcher on; Column IV: the MapReduce Optimization setting with JIT off and the prefetcher on), the microarchitecture-level value improves significantly (Column IV is better than Column III). In Configurations Four and Five, we compare the MapReduce Optimization settings with JIT off and the prefetcher on or off (Column IV: the MapReduce Optimization setting with JIT off and prefetcher on; Column V: the MapReduce Optimization setting with JIT off and prefetcher off). The microarchitecture-level value increases slightly (Column IV is better than Column V). Therefore, using WPC, we can quantitatively investigate the impact of different components on the bottleneck and choose the best configuration.

\subsection{Analyzing and eliminating high L1I Cache Misses of the scale-out Hadoop workloads}~\label{EvaluationsScaleOut}

Our observations corroborated with the previous ones~\cite{ferdman2011clearing} and other influential works~\cite{10.1145/2749469.2750392,7920850}, scale-out Hadoop workloads have higher L1I misses at the typical X86-64 microarchitecture. In this section, we demonstrate how to analyze and eliminate  high L1I Cache Misses of the scale-out Hadoop workloads using WPC, and the microarchitecture platform is Intel Xeon Gold 5120T.

\subsubsection{Break down analysis of the L1I Cache Misses of the Bayes-Hadoop workload}~\label{EvaluationsMO461-1}

Fig.~\ref{method1fig301Intro}-b shows the normalized impact factors of Bayes-Hadoop’s instruction locality. In this Figure, the first layer includes three categories: IR-dependent, ISA-dependent, and microarchitecture-dependent. 
Bayes-Hadoop is implemented with the MapReduce programming framework in Java. The second layer of the IR-dependent category includes the Java language and the MapReduce programming framework. By identifying the MapReduce-related Class name,  WPC can decouple the MapReduce framework bytecodes from the Java execution bytecodes at the IR level. The second layer of the ISA-dependent category includes the Hadoop daemons (HDFS and Yarn schedule daemons) and the language runtime environment (JVM). WPC can decouple the impact of the Hadoop daemons by running the scale-out Hadoop workloads under the standalone mode (the standalone mode would not boot the related Hadoop daemons). We do not change ISAs in this evaluation, so the normalized impact factors of different ISAs are not calculated. The second layer of the
microarchitecture-dependent category includes the OS noise and the specific microarchitecture. We take the OS system call as the OS noise, as they do not run in the user mode. From Fig.~\ref{method1fig301Intro}-b, we can see that the normalized impact factor of Java on the instruction locality is 0.22, and the value of the MapReduce programming framework, JVM,  Hadoop daemons, and specific microarchitecture is 0.14, 0.045, 0.005, and 0.59, respectively. As the normalized impact factor of the kernel mode is not more than 0.001, we consider the OS noise is zero. The other scale-out Hadoop workloads have similar results.

\subsubsection{Optimizations across stacks}~\label{EvaluationsMO461-2}

\begin{figure}[h]
\centering
\includegraphics[scale=1]{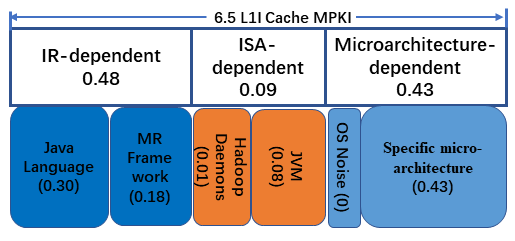}
\caption{Break down the normalized impact factors of the critical components on the L1I Cache Misses of Bayes-Hadoop of optimizations across the stacks on Intel Xeon Gold 5120T.}
\label{FrontEndOpt1}
\end{figure}

We illustrate the impacts of optimizations across the stacks, described in Section~\ref{EvaluationsCrossStackTune}. Fig.~\ref{FrontEndOpt1} shows the breakdown of the instruction locality  of the best configuration (Configuration Four) of the Bayes-Hadoop workload in terms of the normalized impact factors. As the instruction locality improves  by 2.6 times, the increase in the percentage does not imply a worsening situation. For better clarity, we multiply the normalized impact fact with the L1I Cache MPKI of Configuration Four. The MPKI of the MapReduce programming framework changes from 2.4 (16.9*0.14) to 1.2 (6.5*0.18), the MPKI of the JVM changes from 0.76 to 0.52, and the MPKI of the microarchitecture changes from 9.9 to 2.8.

\subsubsection{The optimizations with the MPI implementation}~\label{EvaluationsMO461-3}

\begin{figure}[h]
\centering
\includegraphics[scale=1]{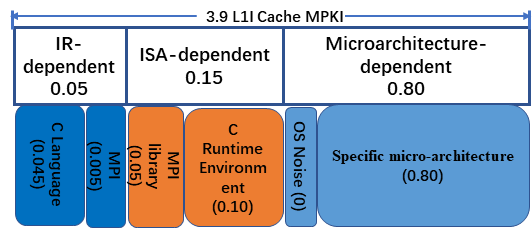}
\caption{Break down the normalized impact factors of the critical components on the L1I Cache Misses of Bayes-MPI on Intel Xeon Gold 5120T.}
\label{FrontEndOpt2}
\end{figure}

The language, programming framework, and runtime environment significantly impact the instruction locality values (0.41). So, we try to change the programming framework from Hadoop to  MPI, another scale-out framework. The language also changes from Java to C. Compared to Hadoop, MPI can dramatically improve the instruction locality. The L1I Cache MPKI of the Bayes-MPI workload decreases by 4.3 times. Fig.~\ref{FrontEndOpt2} shows the normalized impact factors of Bayes-MPI's instruction locality. We can see that the normalized impact factor of the language and programming framework decreases from 0.41 (Java, Hadoop, and JVM) to 0.2 (C, MPI, C runtime environment).

\section{Related Work}

Previous workload characterization methodologies usually focus on only a single level, considering the characterization cost and target ~\cite{6557175,panda2018wait,parsec,bienia2008parsec1,524546,hoste2006comparing,ferdman2011clearing,yasin2014deep,hibench}.


A common methodology is profiling a specific microarchitecture using hardware performance counters~\cite{panda2018wait,ferdman2011clearing,yasin2014deep}. Panda et al. ~\cite{panda2018wait} conducted architecture-dependent workload characterization on SPEC CPU 2017 under seven different ISAs' commercial processor platforms. Then they obtained a representative subset of SPEC CPU 2017 through hierarchical clustering. Yasin ~\cite{yasin2014deep} proposed a top-down performance analysis method to uncover pipeline bottlenecks of out-of-order execution processors based on typical Intel Xeon processors.
Ferdman et al.~\cite{ferdman2011clearing} performed the microarchitecture-dependent workload characterization for the Scale-out workload with the Intel VTune tool on a typical Intel Xeon processor platform. They found that the inefficiency comes from the mismatch between the workload needs and modern processors.

Architecture-independent workload characterization~\cite{parsec,524546,hoste2006comparing} is to profile ISA-instruction traces not tied with specific microarchitecture. Hoste et al. ~\cite{hoste2006comparing} pointed out that there are flaws in comparing the similarity of benchmark programs with workload characteristics that depend on the microarchitecture. They proposed a microarchitecture-independent analysis method based on binary instrumentation and analyzed  122 benchmark programs from six benchmark suites. Biena et al. ~\cite{parsec} adopted Pin-based multiprocessor cache simulator CMP\$im to analyze the PARSEC benchmark's microarchitecture characteristics, including parallelism, locality, computing communication ratio, off-chip memory access, and working set. Woo, et al.~\cite{524546} used a cache-coherent shared address space multiprocessor simulator to analyze SPLASH-2 workloads, including concurrency and load balancing, working sets, communication to computation ratio, and spatial locality.


ISA-independent workload characterization~\cite{6557175} is to leverage the ISA independent characteristic of a compiler intermediate representation (IR). Shao et al. ~\cite{6557175} first proposed an ISA-independent workload analysis method. Based on the intermediate representation of ILJDIT~\cite{campanoni2010highly}, they analyzed SPEC CPU 2000's instruction mix, memory footprint, address entropy, branch entropy, and other instruction-independent workload metrics. By comparing the analysis method based on the X86 binary instruction stream, they found that many traditional workload characteristics are strongly related to the ISA.
Anghel et al. ~\cite{anghel2015instrumentation,anghel2016instrumentation} proposed a platform-independent analysis method PISA. PISA is a modular software analysis framework based on LLVM. They used PISA to analyze the hardware-independent characteristics of the graph analysis benchmark Graph500 and SPEC CPU 2006.


Hsia et al.~\cite{9251259} performed workload characterization on eight deep recommendation models at three levels of the execution stack: algorithms and software, systems platforms, and hardware microarchitecture. They explored the impact of system deployment choices like CPU/GPU and the batch size setting on performance. Furthermore, they analyzed the CPU frontend and backend microarchitectural inefficiencies.


\section{Conclusion}

This article raises an important and challenging workload characterization issue: Can we propose a systematic methodology to uncover each critical component across the stacks contributing what percentages to any specific bottleneck? To answer the above issue, We propose a whole-picture workload characterization (WPC) methodology -- an iterative ORFE loop consisting of Observation, Reference, Fusion, and Exploration -- to facilitate the co-design of software and hardware. As the first step, this article focuses on pipeline efficiency. We build and open-source the WPC tool. Our evaluations show WPC can reveal the impacts of critical components like language, programming framework, runtime environment, ISA, OS, and microarchitecture on the primary pipeline efficiency in a quantitative manner, which can not be done using state-of-the-art and state-of-the-practice tools.

\input{WPC-Corr.bbl}



\end{document}